\documentclass[conference]{IEEEtran}
\IEEEoverridecommandlockouts
\usepackage[english]{babel}
\usepackage{amsthm}
\usepackage{titlesec}
\usepackage[ruled,vlined]{algorithm2e}
\usepackage{algcompatible}
\usepackage{graphicx}
 
\usepackage{float}
\usepackage{ragged2e}
\usepackage[labelfont=bf,font=small,justification=justified,singlelinecheck=false]{caption}
\usepackage{etoolbox}
\usepackage{subcaption}
\makeatletter
\patchcmd{\@makecaption}{\centering}{\justifying}{}{}
\makeatother
\usepackage{tikz}
\usepackage{array}
\usepackage{changepage}
\usepackage[utf8]{inputenc}
\usepackage{pgfplots} 
\usepackage{pgfgantt}
\usepackage{pdflscape}
 \usepackage{relsize}
\usepackage[export]{adjustbox}
\pgfplotsset{compat=newest} 
\pgfplotsset{plot coordinates/math parser=false}
\pgfplotsset{compat=1.18}
\usepackage{pgfplots}
\usetikzlibrary{spy}

\usepackage{cite}
\usepackage{amsmath,amssymb,amsfonts}
\usepackage{dsfont} 
\usepackage{stfloats}
\newtheorem{innercustomgeneric}{\customgenericname}
\providecommand{\customgenericname}{}
\newcommand{\newcustomtheorem}[2]{%
  \newenvironment{#1}[1]
  {%
   \renewcommand\customgenericname{#2}%
   \renewcommand\theinnercustomgeneric{##1}%
   \innercustomgeneric
  }
  {\endinnercustomgeneric}
}

\newcustomtheorem{customthm}{Theorem}
\newcustomtheorem{customlemma}{Lemma}
\newcustomtheorem{customprop}{Proposition}
\newcustomtheorem{customcor}{Corollary}
\usepackage{lipsum}
\usepackage{amsmath}
\usepackage[nolist,printonlyused]{acronym} 
\usepackage{amssymb}
\usepackage{mathtools}
\usepackage{url}
\usepackage{graphicx}  
\usepackage{float}  

\newcommand{\yy}{\mathbf{y}}
\newcommand{\xx}{\mathbf{x}}

\newcommand{\zz}{\mathbf{z}}
\newcommand{\hh}{\mathbf{h}}
\newcommand{\nn}{\mathbf{n}}

\newcommand{\pp}{\mathbf{p}}
\newcommand{\qq}{\mathbf{q}}
\newcommand{\XXi}{\boldsymbol{\Xi}}

\newcommand{\rmr}{{\textrm{r}}}

\newcommand{\blkdiag}{\textrm{blkdiag}}

\newcommand{\Mbs}{{M_{\rm{BS}}}}
\newcommand{\Mbstx}{{M_{\rm{BS}}^{\textrm{Tx}}}}
\newcommand{\Mbsrx}{{M_{\rm{BS}}^{\textrm{Rx}}}}
\newcommand{\Mue}{M_{\rm{UE}}}
\newcommand{\veccs}[1]{ {\rm{vec}}\big(#1\big)  }
\newcommand{\bs}{{\normalfont{\textrm{BS}}}}
\newcommand{\ue}{{\normalfont{\textrm{UE}}}}

\newcommand{\rmd}{{\textrm{D}}}

\newcommand{\oomega}{\boldsymbol{\omega}}

\newcommand{\rrho}{\boldsymbol{\rho}}

\newcommand{\rmp}{{\textrm{p}}}

\newcommand{\BB}{\mathbf{B}}

\newcommand{\GG}{\mathbf{G}}
\newcommand{\YY}{\mathbf{Y}}
\newcommand{\NN}{\mathbf{N}}
\newcommand{\HH}{\mathbf{H}}

\newcommand{\WW}{\mathbf{W}}

\newcommand{\FF}{\mathbf{F}}
\newcommand{\PP}{\mathbf{P}}

\newcommand{\RR}{\mathbf{R}}
\newcommand{\DD}{\mathbf{D}}
\newcommand{\CC}{\mathbf{C}}

\newcommand{\diag}{\text{diag}}

\newcommand{\hermit}{\mathsf{H}}

\usepackage{accents}

\newcommand{\JJ}{\mathbf{J}}
\newcommand{\TT}{\mathbf{T}}

\newcommand{\MM}{\mathbf{M}}

\newcommand{\xxb}{\boldsymbol{x}}

\newcommand{\yyb}{\boldsymbol{y}}
\newcommand{\YYb}{\boldsymbol{Y}}
\newcommand{\ttau}{\boldsymbol{\tau}}
\newcommand{\mmu}{\boldsymbol{\mu}}

\newcommand{\ptot}{{p_\textrm{tot}}}
\newcommand{\AAb}{\mathbf{A}}

\newcommand{\norm}[1]{\left\lVert#1\right\rVert}

\newcommand{\rmtx}{{\rm{Tx}}}
\newcommand{\rmrx}{{\rm{Rx}}}

\newcommand{\ee}{\mathbf{e}}

\newcommand{\ttheta}{\boldsymbol{\theta}}

\newcommand{\vecc}{\text{vec}}

\newcommand{\EE}{\mathbf{E}}

\newcommand{\eeta}{\boldsymbol{\eta}}

\newcommand{\arx}{\mathbf{a}_\rmrx}
\newcommand{\aue}{\mathbf{a}_{\textrm{UE}}}
\newcommand{\dl}{ {\textrm{DL}}}
\newcommand{\ul}{ {\textrm{UL}}}
\newcommand{\atx}{\mathbf{a}_\rmtx}

\newcommand{\tx}{{\textrm{Tx}}}

\newcommand{\cc}{\mathbf{c}}

\acrodef{SISO}[SISO]{single-input single-output}
\acrodef{PRB}[PRB]{physical resource block}
\acrodef{AP}[AP]{access point}
\acrodef{AR}{autoregressive}
\acrodef{UE}[UE]{user equipment}
\acrodef{ULA}[ULA]{uniform linear array}
\acrodef{ML}[ML]{maximum likelihood}
\acrodef{CPU}[CPU]{central processing unit}
\acrodef{DL}{downlink}
\acrodef{UL}{uplink}
\acrodef{FPP}[FPP]{Feasible-point pursuit}
\acrodef{LoS}[LoS]{line of sight}
\acrodef{NLoS}[NLoS]{non-line-of-sight}
\acrodef{RCS}[RCS]{radar cross section}
\acrodef{AoD}[AoD]{angle of departure}
\acrodef{AoA}[AoA]{angle of arrival}
\acrodef{CRB}[CRB]{Cramer-Rao bound}
\acrodef{FIM}[FIM]{Fisher information matrix}
\acrodef{AN}[AN]{artificial noise}
\acrodef{SINR}[SINR]{signal-to-interference-plus-noise ratio}
\acrodef{TDD}[TDD]{time-division duplexing}
\acrodef{SNR}[SNR]{signal-to-noise ratio}
\acrodef{QoS}[QoS]{quality of service}
\acrodef{SDR}[SDR]{semi-definite relaxation}
\acrodef{SDP}[SDP]{semi-definite program}
\acrodef{ISAC}[ISAC]{integrated sensing and communications}
\acrodef{PLS}[PLS]{physical layer security}
\acrodef{SIC}[SIC]{successive interference cancellation}
\acrodef{CSI}[CSI]{channel state information}
\acrodef{CSIT}[CSIT]{channel state information at the transmitter}
\acrodef{CSIR}[CSIR]{channel state information at the receiver}
\acrodef{MUI}[MUI]{multi-user interference}
\acrodef{RIS}[RIS]{Reconfigurable intelligent surface}
\acrodef{AO}[AO]{alternating optimization}
\acrodef{SIMO}[SIMO]{Single Input Multiple Output}
\acrodef{MISO}[MISO]{multiple-intput single output}
\acrodef{MIMO}[MIMO]{multiple-input multiple-output}
\acrodef{MU}{multi-user}
\acrodef{BS}[BS]{base station}
\acrodef{CEE}[CEE]{channel estimation error}
\acrodef{CCP}[CCP]{convex-concave procedure}
\acrodef{MRT}[MRT]{ maximum-ratio transmission}
\acrodef{MRC}[MRC]{ maximum-ratio combining}
\acrodef{MM}[MM]{Minorization-Maximization}
\acrodef{PSD}[PSD]{positive semi-definite}
\acrodef{RZF}[RZF]{Regularized zero forcing}
\acrodef{CRZF}[CRZF]{Centralized regularized zero forcing}
\acrodef{RZF}[RZF]{regularized zero forcing}
\acrodef{LPZF}[LPZF]{Local Partial zero forcing}
\acrodef{LZF}[LZF]{Local zero forcing}
\acrodef{NF}[NF]{Near Field}
\acrodef{FF}[FF]{Far Field}
\acrodef{BD}[BD]{Beyond-diagonal}
\acrodef{OFDM}[OFDM]{orthogonal frequency division multiplexing}
\acrodef{MAPRT}[MAPRT]{maximum a-posteriori ratio test}
\acrodef{LRT}[LRT]{ likelihood ratio test}
\acrodef{CDF}[CDF]{Cumulative distribution function}
\acrodef{UPA}[UPA]{uniform planar array}
\acrodef{LMMSE}[LMMSE]{linear minimum mean square error}
\acrodef{MMSE}[MMSE]{minimum mean square error}
\acrodef{SE}[SE]{spectral efficiency}
\acrodef{CNR}[CNR]{clutter to noise ratio}
\acrodef{SCNR}[SCNR]{signal to clutter and noise ratio}
\acrodef{SOCP}[SOCP]{second order cone program}
\acrodef{TTD}[TTD]{true time delay}
\acrodef{PS}[PS]{phase shifter}
\acrodef{PEB}[PEB]{Position error bound}
\acrodef{JRC}[JRC]{Joint radar and communication}
\acrodef{GP}[GP]{gradient projection}
\acrodef{MO}[MO]{Manifold Optimization}
\acrodef{DEB}[DEB]{direction error bound}
\acrodef{REB}[REB]{range error bound}
\acrodef{LS}[LS]{least-squares}
\acrodef{MUSIC}[MUSIC]{MUltiple SIgnal Classification}
\acrodef{SR}[SR]{sum rate}
\acrodef{WMMSE}[WMMSE]{weighted-minimum mean square error}
\acrodef{BCD}[BCD]{block-coordinate descent}
\acrodef{SOCP}[SOCP]{second-order cone program}
\acrodef{DFT}[DFT]{discrete Fourier transform}
\acrodef{PSO}[PSO]{particle swarm optimization}
\acrodef{SIRV}[SIRV]{spherically-invariant random vector}
\acrodef{MF}[MF]{matched filter}
\acrodef{SCR}[SCR]{signal to clutter ratio}
\acrodef{CFAR}[CFAR]{constant false alarm rate}
\acrodef{RA}[RA]{range-angle}
\acrodef{RV}[RV]{range-velocity}
\acrodef{MU-MIMO}{multi-user multiple input multiple output}
\acrodef{NMSE}{normalized mean squared error}
\acrodef{SV}[SV]{singular value}

\begin{document}

\title{Cramer-Rao Bound Analysis of Bistatic ISAC Under Partial Symbol Knowledge and Clutter}

\author{
Steven~Rivetti$^\dagger$,
G\'abor Fodor$^{*,\dagger}$,
Emil~Bj\"ornson$^\dagger$, 
        Mikael~Skoglund$^\dagger$ \\
        
       {\small$^\dagger$School of Electrical Engineering and Computer Science (EECS),
        KTH Royal Institute of Technology, Sweden.} \\
        
        {\small$^*$ Ericsson Research, Stockholm, Sweden.}
  \thanks{
This work was supported by the SUCCESS project (FUS21-0026) and by the SAICOM project (FUS21-0004), both funded by the Swedish Foundation for Strategic Research (SSF).}

}

\maketitle

\begin{abstract}
Integrated sensing and communication (ISAC) systems rely on communication waveforms to perform sensing tasks, thus making their sensing performance strongly dependent on the level of communication-symbol knowledge available to the sensing receivers. However, the existing literature fails to capture this dependency, often relying on full-symbol-knowledge assumptions.
In this paper, we present a Cramer–Rao bound (CRB) analysis of a bistatic ISAC network with heterogeneous uplink and downlink illumination and structured clutter.
We consider different symbol-knowledge regimes by modeling unknown communication symbols as nuisance parameters. 
Assuming a temporal evolution of the communication channel,
we derive a correlation-aware channel estimator and an expression for the UEs' uplink spectral efficiency. 
Numerical results show the CRB degradation induced by clutter and symbol uncertainty and how this can affect resource allocation policies. We also show the performance gain of our channel estimator relative to conventional block-fading architectures. 



\end{abstract}

\section{introduction}
Emerging applications, such as autonomous driving and \ac{UE} localization, require wireless sixth-generation networks to support both communication and sensing functions. \ac{ISAC} allows for sensing-communication cooperation tasks, greatly improving the system's efficiency \cite{liu2022integrated}.
Distributed and collaborative \ac{ISAC} architectures, in which the physical separation and orchestration between the transceivers allow the system to reduce self-interference and improve target observability,
have recently gained attention \cite{strinati2024distributed}.
However, such gains may come at the expense of an increased hardware cost and the need for inter-\acp{BS} synchronization \cite{masouros2025synch}, which are notoriously difficult to model and evaluate.
Therefore, a fundamental challenge of \ac{ISAC} system design is characterizing realistic performance constraints, including the impact of imperfect \ac{CSI}, clutter, and channel aging.
Perhaps the most important of those is the presence of environmental clutter  \cite{swindlehurst2026clutter}.
In \cite{Vinogradova:13}, a target detector in the presence of clutter is presented, and the probability of detection is shown for different levels of task integration. On the same note, \cite{rivetti2024clutter} characterizes the detection performance of a distributed \ac{MIMO} network affected by clutter.
A dynamically changing propagation environment also induces temporal correlation on the communication links between \acp{UE} and \acp{BS} \cite{Zhang:26}.
This aspect, which block-fading-based models disregard, is crucial for tracking channel aging \cite{fodor2023optimizing}, a phenomenon that describes the channel estimate drift from the actual value after the channel estimation phase.
If a block structure is to be preserved, \cite{Fodor:21} characterizes the performance of a \ac{MMSE} channel estimator that allows different blocks to have non-negligible time correlation.
The use of random communication waveforms as sensing illumination \cite{liu2023deterministic} poses an additional challenge to 
distributed \ac{ISAC} systems.
Indeed, it is crucial to characterize the availability, or the lack thereof, of knowledge on the transmitted communication symbols at the sensing receiver\cite{Lu:26}.
When such knowledge is available, the waveform used for radar illumination is deterministic, while in the absence of such knowledge, the illumination waveform becomes stochastic and random, often with a zero-mean Gaussian distribution \cite[Chapter 12]{Gershman:08}.
An interesting comparison between these models can be found in \cite{fodor2024trade}, where the authors compare the \acp{CRB} on the \ac{AoA} of a single target at a high abstraction level across both waveform regimes.
\begin{figure}[t!]
\begin{center}
   \resizebox{0.4\textwidth}{!}{
     \includegraphics[]{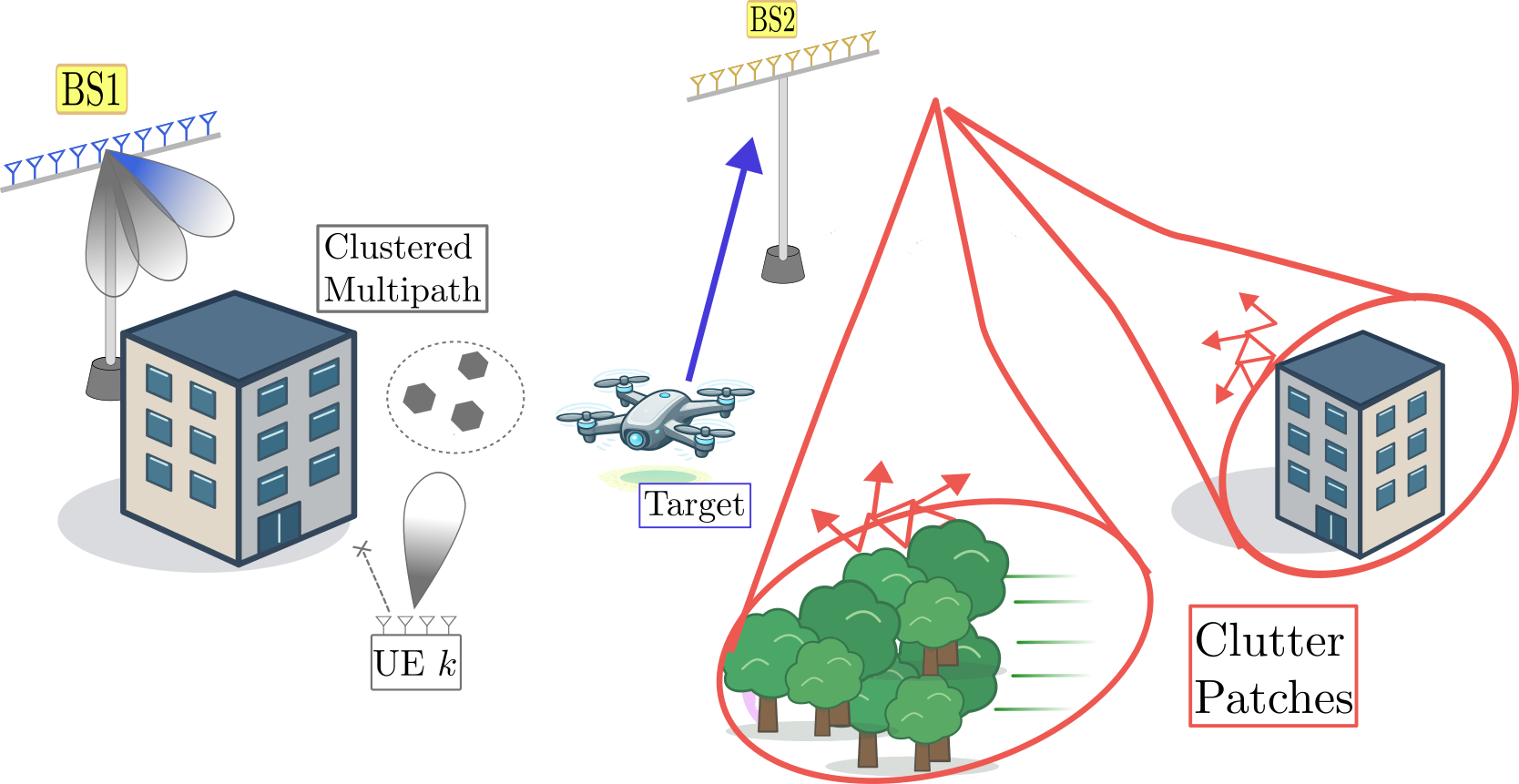}}
      \vspace{-2mm}
	 \caption{Depiction of the considered bistatic ISAC network. \ac{BS}1 transmits the integrated waveform in DL and receives the UE communication data in UL. BS2 is the sensing receiver.  
     The blue arrow indicates the target echoes, while the ellipses indicate clutter patches: clutter is generated by the superimposition of clutter patches with finite angular, Doppler, and delay spreads.
    }
    \label{scenario}
\end{center}
  \vspace*{-6mm}
\end{figure}

In this paper, we want to find out how much imperfect transmit symbol knowledge hinders sensing accuracy. We do this through \ac{CRB} analysis of a bistatic \ac{ISAC} system using \ac{OFDM} waveforms in the presence of structured clutter and imperfect transmit symbol knowledge (Fig. \ref{scenario}).
We consider heterogeneous sensing illumination as the receiving BS2 
receives the sensing echoes in both the \ac{UL} and \ac{DL}. 
We derive the target's \ac{CRB} for different degrees of transmit symbol knowledge: when the latter are unknown, they are treated as nuisance parameters and later marginalized.
On the other hand, we derive an \ac{MMSE} \ac{UL} channel estimator which leverages multiple past pilot observations. The \acp{UE} communication performance are assessed through the derived \ac{UL} \ac{SE}.
The proposed framework assesses how much imperfect receiver-side transmit symbol knowledge and clutter degrade sensing accuracy, and how this loss can be mitigated through resource allocation.
It also shows that pilot design and leveraging channel temporal correlation lead to a substantial SE gain compared to block-fading approaches.

\section{System Model}
\begin{figure}[t!]
\begin{center}
   \resizebox{0.44\textwidth}{!}{
     \includegraphics[]{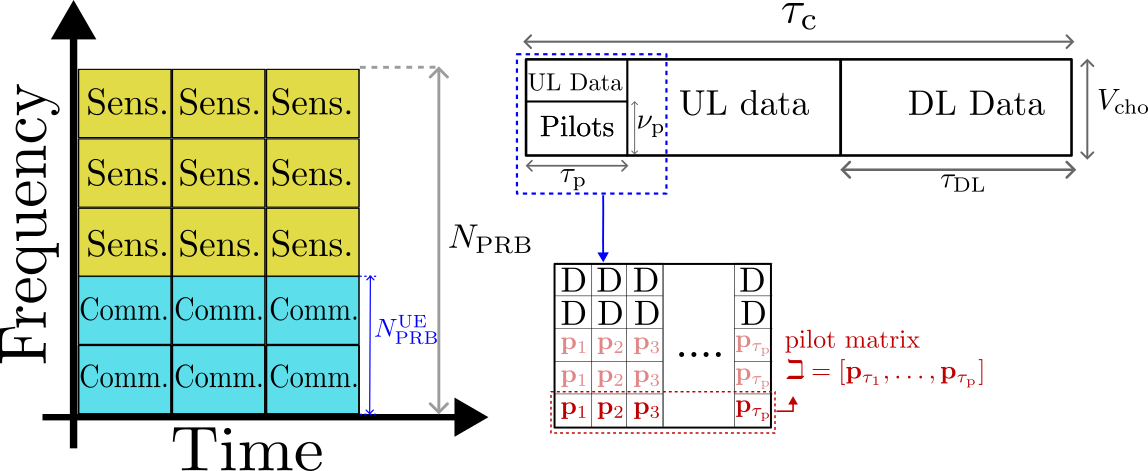}}
      \vspace{-2mm}
	 \caption{BS$1$ resource scheduling scheme and frame structure: $\pp_t$ is the transmitted pilot vector, with the semi-transparent replicas being optional repetitions of the former for \ac{SNR} purposes. The UE index $k$ has been dropped here for ease of notation.}
    \label{prb grid}
\end{center}
  \vspace{-4mm}
\end{figure}

We consider the coordinated bistatic ISAC deployment shown in Fig.~\ref{scenario}, which realizes continuous sensing at every symbol time without disrupting the communication transmission, at the cost of additional synchronization. An application scenario of this deployment can be found in cooperative traffic monitoring along a street canyon.
BS2 receives target echoes and is assumed to know the resource allocation policy and precoders, but not necessarily the transmitted data symbols.
BS1 is equipped with $\Mbstx$ antennas and located at the origin, BS2 has $\Mbsrx$ antennas and is located at $\qq=[q_{x},q_{y}]^\top$. The network serves $K$  communication \acp{UE},  each equipped with $\Mue$ antennas and located at $\ee_k=[e_{k,x},e_{k,y}]^\top$.
The \acp{BS} perform \ac{OFDM} transmission around the carrier frequency $f_c$,  over $V$ active subcarriers with a subcarrier spacing of $\Delta_f$.
The subcarriers are partitioned into $N_\textrm{PRB}$ \acp{PRB} of $V_\textrm{cho}$ subcarriers each.
We define the subcarrier index $v$ and the frequency block index $n$.
The \acp{UE} also employ OFDM transmission over  $N_\textrm{PRB}^\textrm{UE}$ of the available \acp{PRB}, and remain silent over the remaining ones.
The symbol duration is $T=1/\Delta_f + T_\textrm{cp} $, with $T_\textrm{cp}$ being the cyclic prefix duration.
The PRB time duration is  $\tau_\textrm{c}$ OFDM symbols: within each \ac{PRB}, the channel is assumed constant.
Fig. \ref{prb grid} shows BS1's resource scheduling policy: assuming \ac{TDD} and \ac{UL}/\ac{DL} channel reciprocity, UL channel estimates are reused for DL precoding.
The target echoes are used to calculate the position and velocity of a target, located at $\mathbf{t}=[t_{x},t_{y}]^\top$.
We define the symbol time index $i$, with a time unit $T$, and the block time index $b$, with a time unit of $\tau_\textrm{c}T$.
The channel between \ac{BS}1 and \ac{UE} $k$ at block $\{b,n\}$ is denoted by $\HH_{k,b,n} \in \mathds{C}^{\Mbstx \times \Mue}$. 
Let us build $\hh_{k,b,n}=\vecc(\HH_{k,b,n})\in \mathbb{C}^{\Mbstx\Mue}$,  
which follows a correlated Rayleigh fading distribution, with a Kronecker covariance model:
$\hh_{k,b,n} \sim \mathcal{CN}(\mathbf{0},\CC_k)$, where 
\begin{align}
\vspace{-3mm}
\CC_k=\CC_{\textrm{UE},k} \otimes\CC_{\textrm{BS},k} \in \mathds{C}^{\Mbstx\Mue \times \Mbstx\Mue}.
\end{align}
Here $\CC_{\textrm{BS},k} \in \mathds{C}^{\Mbs \times \Mbs}$, 
and $\CC_{\textrm{UE},k} \in \mathds{C}^{\Mue \times \Mue}$ are the BS and UE side correlation matrices respectively.
We consider the presence of non-negligible time correlation between blocks, while, for simplicity, we neglect the frequency correlation. 
Time correlation is modeled as a first-order \ac{AR} process with a scaled identity state-transition matrix \cite{fodor2023optimizing}.
The correlation between $\hh_{k,b,n}$ and $\hh_{k,b',n}$ is thus defined as
   $ \mathds{E}[\hh_{k,b,n}\hh^\hermit_{k,b',n}]=\CC_k\zeta_k(|b-b'|),$
where $\zeta_k(|b-b'|)$ captures the correlation decay. It is chosen as $\zeta_k(|b-b'|)=J_0(2\pi \tau_\textrm{c}T  |b-b'|\norm{\oomega_k}/\lambda_c)$, where $\oomega_k$ is UE $k$'s velocity vector and $J_0$ is the zero-th order Bessel function\cite{wang2003efficient}.
The Bessel correlation decay is adopted as an isotropic aging benchmark, which averages the Doppler shift over uniformly distributed \acp{AoA}.
The sensing channel takes different shapes based on the frame's phase: $\GG_{\textrm{BS},i,v} \in \mathds{C}^{\Mbsrx \times \Mbstx}$ is the frequency-domain channel accounting for the BS1-Target-BS2 propagation path, while $\GG_{\textrm{UE},k,i,v} \in \mathds{C}^{\Mbsrx \times \Mue}$ models the UE $k$-Target-BS2 path.
These channels are modeled as 
\begin{align}
    &\GG_{\textrm{BS},i,v} \triangleq \beta_\textrm{BS} e^{-j2\pi f_{\textrm{D}}Ti }e^{j2\pi\Delta_f v\tau_\textrm{BS}}\arx(\psi)\atx(\theta_\textrm{BS})^\hermit\\
    &\GG_{\textrm{UE},k,i,v}\hspace{-1mm}\triangleq \beta_{\textrm{UE},k}e^{-j2\pi f_{\textrm{D},k}Ti }e^{j2\pi\Delta_f v\tau_{k}}\arx(\psi)\aue(\theta_{k})^\hermit
\end{align}
where 
\begin{align}
    &\psi=\textrm{atan}2(q_{y}-t_{y},q_{x}-t_{x}),~\theta_\textrm{BS}=\textrm{atan}2(t_y,t_x),\\
    &\theta_{\textrm{UE},k}=-\textrm{atan}2(p_{y,k},p_{x,k}),~\tau_{k}=(\norm{\mathbf{t}-\ee_k}+ \norm{\mathbf{q}-\mathbf{t}})/c \nonumber\\
    &\tau_\textrm{BS}=(\norm{\mathbf{t}}+ \norm{\mathbf{q}-\mathbf{t}})/c,\nonumber
\end{align}
and $c$ is the speed of light.
Here $\atx(\cdot)\in \mathbb{C}^{\Mbstx}$, $\arx(\cdot)\in \mathbb{C}^{\Mbsrx}$. and $\aue(\cdot)\in \mathbb{C}^{\Mue}$ are the array response vectors of the \acp{BS} and the \acp{UE}, respectively.
The Doppler shifts of the target w.r.t. the BS and UEs are defined as 
\begin{align}
    &f_\rmd=\frac{1}{\lambda_c}\left[ \oomega^\top \frac{\mathbf{t}}{\norm{\mathbf{t}}} - \oomega^\top \frac{\mathbf{q}-\mathbf{t}}{\norm{\mathbf{q}-\mathbf{t}}} \right]\\
    &f_{\rmd,k}=\frac{1}{\lambda_c}\left[ (\oomega-\oomega_k)^\top \frac{\mathbf{t}-\ee_k}{\norm{\mathbf{t}-\ee_k}} - \oomega^\top \frac{\mathbf{q}-\mathbf{t}}{\norm{\mathbf{q}-\mathbf{t}}}  \right],
\end{align}
where $\oomega$ and $\oomega_k$ are the target and UE $k$ velocity vectors, respectively. 
Here $\beta_{\textrm{UE},k},~\beta_{\textrm{BS}}$ are the complex channel gains of the target, both defined as
$\beta_{\textrm{UE},k},\,\beta_{\textrm{BS}}\sim\mathcal{CN}\left(0,\sqrt{ c^2 \delta_{\textrm{tg}}^2 /(4\pi)^3f_c^2r^4}  \right)$, with $r=c\tau_\textrm{BS}$ or $r=c\tau_k$.
Let us assume that the DL symbol time $i$ and subcarrier $v$ belong to a PRB  with time-frequency indexes $\{b,n\}$. Then,
the frequency-domain observation at UE $k$ is 
\begin{align}
\label{y generic data}
\yy_{i,v,k}
&\triangleq
\alpha_k\HH_{k,b,n}^\hermit\FF_{i,v}\PP_{i,v}\xx_{i,v} + \nn_{i,v,k},
\end{align}
\newpage
\vspace*{0.03in}
where $\PP_{i,v}=\diag(\boldsymbol{\rho}_{i,v})$, $\FF_{i,v} \in \mathds{C}^{\Mbs \times S}$ is the precoder matrix, and $\boldsymbol{\rho}_{i,v}\triangleq \hspace{-1mm}\left[\sqrt{\rho_{i,v,1}},\dots,\sqrt{\rho_{i,v,S}}\right]^\top$ represents the power allocation policy.
Each column of $\FF_{i,v}$ is normalized to have a unitary norm. 
 The transmitted symbols  
are $\xx_{i,v}=[x_{i,v,1},\dots,x_{i,v,S}]^\top$, with $x_{i,v,s}\sim \mathcal{CN}(0,1)$.
Lastly, $\nn_{i,v,k}\sim \mathcal{CN}(0,\mathbf{I}_{\Mue}\sigma_{k}^2)$ is UE $k$ receiver noise.
The UL signal observed at  BS1 is defined as 
\vspace{-1mm}
\begin{align}
\yy_{i,v}\triangleq\sum_{k=1}^K\alpha_k\HH_{k,b,n}\WW_{k} \boldsymbol{P}_{k}\boldsymbol{x}_{k,i,v}  + \nn_{i,v},
\end{align}
where $\boldsymbol{P}_{k }=\diag(\boldsymbol{\varrho}_{k })$.
Here $\WW_{k} \in \mathbb{C}^{\Mue \times \Mue}, \boldsymbol{\varrho}_{k},~\boldsymbol{x}_{k,i,v}$ are the precoding matrix, the power allocation policy, and the transmitted symbols adopted by UE $k$ in its UL transmission.
Due to the lack of \ac{CSI} at the UEs, the same $\WW_k$ is applied to all of their subcarriers and symbol times.
We assume that  $\boldsymbol{x}_{k,i,v}\sim \mathcal{CN}(\mathbf{0},\mathbf{I}_{\Mue} )$. The BS receiver noise is $\nn_{i,v}\sim \mathcal{CN}(0,\mathbf{I}_{\Mbstx}\sigma_{\bs}^2)$.
Let us define the sets $\mathcal{U}$ and $\mathcal{D}$ that contain all the $\{i,v\}$ bins where, respectively, UL and DL communication takes place.
The observed radar echoes at BS2 are then defined as 
\begin{align}\label{y radar}
\boldsymbol{y}_{i,v}\triangleq\begin{cases}
\boldsymbol{y}^{\textrm{UL}}_{i,v}\triangleq
\mmu_{\ul,i,v}+\cc_{\textrm{UL},i,v}\hspace{-1mm} +\nn_{v,i}^\rmr & \{i,v\} \in \mathcal{U}, \\
\boldsymbol{y}^{\textrm{DL}}_{i,v}\triangleq \mmu_{\dl,i,v}+ \cc_{\dl,i,v}+\nn_{v,i}^\rmr & \{i,v\} \in \mathcal{D}, \\
\end{cases}
\vspace*{-4mm}
\end{align}
where 
\vspace*{-2mm}
\begin{align}
&\mmu_{\dl,i,v}=\GG_{\textrm{BS},i,v} \FF_{i,v}\PP_{i,v}\xx_{i,v}\\
&\mmu_{\ul,i,v}=\sum\nolimits_{k=1}^K\GG_{\textrm{UE},k,i,v}\WW_{k,i,v}\boldsymbol{P}_{k,i,v}\boldsymbol{x}_{k,i,v}.
\end{align}
Once again, $\nn_{i,v}^\rmr\sim \mathcal{CN}(\mathbf{0},\sigma_\bs^2\mathbf{I}_{\Mbsrx} )$ is the receiver noise.
We consider the presence of clutter affecting the bistatic ISAC operations:
let $\cc_\textrm{DL}\in\mathbb{C}^{\Mbsrx I_\textrm{DL} V \times \Mbsrx I_\textrm{DL} V}$ and 
$\cc_\textrm{UL}\in\mathbb{C}^{\Mbsrx I_\textrm{UL} V_\textrm{UE} \times \Mbsrx I_\textrm{UL} V_\textrm{UE}}$
denote the stacked DL and UL clutter vectors, respectively.
Here $I_\textrm{DL}$ and $I_\textrm{UL}$ are the total number of UL and DL symbol times, respectively, while $V_\textrm{UE}=N_\textrm{PRB}^\textrm{UE}V_\textrm{cho}$. 
These vectors are $\cc_\textrm{DL}\sim\mathcal{CN}(\mathbf{0},\delta_\textrm{cl}^2\BB_\dl)$, and $\cc_\textrm{UL}\sim\mathcal{CN}(\mathbf{0},\kappa_\textrm{cl}^2\delta_\textrm{cl}^2\BB_\ul)$ with $\delta_\textrm{cl}^2$ being  the clutter's texture.
We adopt the Kronecker-separable model: Here clutter arises for a summation of approximately uncorrelated clutter patches, allowing for  the separation of the spatial, temporal, and frequency components of the total covariance matrix. 
The adopted patched model captures spatially extended clutter with finite angular, Doppler, and delay spreads, 
while deterministic specular reflections are outside the present scope.
Here $\kappa_\textrm{cl}^2 \leq 1$ models the reduced clutter texture during UL illumination, originating from the reduced UE transmit power.

\section{Channel Estimation and Spectral Efficiency}
Fig.~\ref{prb grid} shows how UE $k$ transmits its pilot matrix $\beth_k=[\pp_{k,\tau_1},\dots,\pp_{k,\tau_\textrm{p}}] \in \mathbb{C}^{\Mue \times \tau_\rmp }$ over $\tau_\rmp$ symbol times, repeating it over $\nu_\rmp$ subcarriers to boost the \ac{SNR}. Data symbols are scheduled on
the remaining $V_\textrm{cho}-\nu_p$ subcarriers.
The pilot matrices are mutually orthogonal.
The following considerations apply to any \ac{PRB}; we thus omit the index $n$.
The received pilot matrix on the first subcarrier, denoted with $1$, is
\begin{align}
    \YY_{b,1,k} = \sum\nolimits_{k=1}^K \alpha_k\HH_{b,k}\WW_{k}\boldsymbol{P}_k\boldsymbol{\beth}_k
    + \NN\in \mathbb{C}^{\Mbstx \times \tau_\rmp},
\end{align}
where $\NN$ represents the receiver noise matrix, with entries $[\NN]_{l,m}\sim\mathcal{CN}(0,\sigma_k^2)$.
This observation is then de-spread $\widetilde{\YY}_{b,1,k}\hspace{-1mm}\triangleq\YY_{b,1,k}\boldsymbol{\beth}_k^\hermit/\sqrt{\tau_\rmp}$ .
The BS then collects observation on all the $\nu_\rmp$ subcarriers in $\widetilde{\yy}_{b,k}=[\veccs{\widetilde{\YY}_{b,1,k}}^\top,\dots,\veccs{\widetilde{\YY}_{b,\nu_\rmp,k}}^\top]$: this can be written as 
\begin{align}
\widetilde{\yy}_{b,k}=\hspace{-1mm}
\alpha_k\sqrt{\tau_\rmp} \DD_k \hh_{k,b} +  \vecc(\widetilde{\NN})\in \mathbb{C}^{\Mbstx\Mue  \nu_\rmp }
\end{align}
where $\DD_k=\mathbf{1}_{\nu_\rmp} \otimes(\WW_{k}^0\boldsymbol{P}_k^0)^\top \otimes \mathbf{I}_{\Mbstx}$.
The receiver estimates $\HH_{k,b}$ using $p$ pilot observations in addition to the present one, for a total of $p_\textrm{tot}=p+1$ observations, collected in $\widetilde{\yy}^\ptot_{k}\hspace{-1mm}=
    [ \widetilde{\yy}^{\top}_{b,k}\dots\widetilde{\yy}^{\top}_{b-p,k}]^\top\hspace{-1mm}\in \mathbb{C}^{\Mbstx\Mue  \nu_\rmp  \ptot}.$
The \ac{MMSE} channel estimate of $\hh_{b,k}$ is \cite[Lemma 1]{fodor2023optimizing}
    \begin{align}
    &\widehat{\hh}_{k,b} = \alpha_k \sqrt{\tau_\rmp}\hspace{1mm}\EE_{k}\DD_{k,\ptot}^\hermit\hspace{0.5mm}\overline{\AAb}_{k}^{-1}\hspace{0.5mm}\widetilde{\yy}^\ptot_{k} ,
     \label{hmmse}
     \end{align}
where
\vspace*{-2mm}
\begin{align}
&\overline{\AAb}_{k}=\alpha_k^2\tau_\rmp\DD_{k,\ptot}\MM_k\DD_{k,\ptot}^\hermit + \sigma^2_k\mathbf{I}_{\ptot\Mue\Mbstx\nu_\rmp}\\
&\DD_{k,\ptot} = \mathbf{I}_{\ptot} \otimes \DD_k,\quad \EE_{k}=\boldsymbol{\zeta}_p \otimes \CC_k \\
&\MM_{k}= \mathcal{T}\left(\boldsymbol{\zeta}_p\right) \otimes \CC_k,\quad\boldsymbol{\zeta}_p=\left[\zeta_k(0),\dots,\zeta_k(p) \right].
\end{align}  
Here  $\mathcal{T}(\boldsymbol{\zeta}_p)$ denotes the Toeplitz matrix generated by $\boldsymbol{\zeta}_p$.The channel estimate $\widehat{\hh}_{k,b}$ is 
\vspace*{-2mm}
\begin{align}
\widehat{\hh}_{k,b}\sim \mathcal{CN}(\mathbf{0},\widehat{\boldsymbol{\Xi}}_{k,b}= \alpha_k\tau_\rmp\EE_{k}\DD_{k,\ptot}^\hermit \overline{\AAb}_{k}^{-1}\DD_{k,\ptot}\EE_{k}^{\hermit}).  
\end{align}
The channel estimation error $\widetilde{\hh}_{k,b}=\widehat{\hh}_{k,b}-\hh_{k,b}$ is then $\widetilde{\hh}_{k,b}\sim \mathcal{CN}(\mathbf{0},\widetilde{\boldsymbol{\Xi}}_{k,b}=\CC_k-\widehat{\XXi}_{k,b})$ .  
We now derive an achievable \ac{UL} \ac{SE}, assuming imperfect CSI at BS1 and no CSI at the \acp{UE}. 
The following considerations apply to any PRB with block time $b$; we thus omit the indices $i,v,$and $n$.
Under the assumption of successive interference cancellation and by defining $\TT_k$ as an \ac{MMSE} combining matrix\cite{Fodor:21}, an achievable \ac{SE}\cite{debbah2022uplink} on subcarrier $v$, belonging to a PRB with frequency index $n$, for UE $k$ is given in \eqref{multipage} on the top of the next page, where $\overline{\WW}_{k}=\WW_{k}\boldsymbol{P}_{k}$. Here $\widehat{\HH}_{k,b}\, \textrm{and}\, \widetilde{\HH}_{k,b}$
 have been obtained by reshaping $\widehat{\hh}_{k,b}$
and $\widetilde{\hh}_{k,b}$.
The pre-log factor is represented by $\beta_v$: If the subcarrier $v$ has been used for pilot transmission, then $\beta_v=1-\frac{\tau_{\rmp}+\tau_\textrm{DL}}{\tau_c}$, otherwise $\beta_v=1-\frac{\tau_\textrm{DL}}{\tau_c}$.

\begin{figure*}[t!]
  \begin{align}\label{multipage}
&\hspace{-3mm}{\rm{SE}}_{k,v}\hspace{-1mm}= \beta_v\mathbb{E}\left\{ \log_2 \left\vert 
   \mathbf{I}_{\Mue} \hspace{-1mm}+ \alpha_k^2\overline{\WW}_k^\hermit \widehat{\HH}_{k,b,n}^\hermit \hspace{-1mm} 
   \left(\sum_{\substack{l=1,\\ l \neq k}}^K \hspace{-1mm}\alpha_l^2\widehat{\HH}_{l,b,n} \overline{\WW}_l\overline{\WW}_l^\hermit\widehat{\HH}_{l,b,n}^\hermit + 
   \sum_{l=1}^K \hspace{-1mm}\alpha_l^2  \textrm{Cov}[\widetilde{\HH}_{l,b,n}\WW_{l}] +\sigma^2\mathbf{I}_{\Mbstx}\hspace{-1.5mm} \right)^{-1}\hspace{-4mm}\widehat{\HH}_{k,b,n} \overline{\WW}_k
   \right\vert           
   \right\}
\end{align} 
\hrulefill
\end{figure*}

\section{Cramer-Rao Bound Derivation}
A key issue in \ac{FIM} modeling is the availability of transmit-symbol knowledge to the sensing receiver.
By symbol knowledge, we indicate the availability at BS2 of the exact complex-valued transmitted symbols: If such knowledge is not available, said symbols cannot be removed from the sensing observation through matched filtering, which inevitably degrades the estimation process. The unknown symbols are here modeled as deterministic nuisance parameters.
The precoders employed by BS1 and the UEs are assumed to be known by BS2, which is reasonable as precoders change at most once per PRB.
We characterize three operating regimes: clairvoyant bound, where transmit symbols are always known, fully unknown bound, where the transmitted symbols are never known, and a hybrid regime. 
The target's vector of deterministic unknown parameters is 
\vspace*{-0.5mm}
\begin{align}
    \eeta=[\ttheta^\top,\psi,\ttau^\top,\oomega^\top,\Re \{\boldsymbol{\beta}^\top\},\Im\{\boldsymbol{\beta}^\top\}]^\top\in \mathbb{R}^{4(K+1)+3},
\end{align} 
where $\ttheta=[\theta_\textrm{BS},\theta_1,\dots,\theta_K]^\top$, $\ttau=[\tau_\textrm{BS},\tau_1,\dots,\tau_K]^\top$, $\boldsymbol{\beta}=[\beta_\textrm{BS},\beta_1,\dots,\beta_K]^\top$.
We construct the UL and DL data cubes as $\boldsymbol{Y}^\dl \in \mathds{C}^{\Mbsrx \times I_\dl \times V}$ and $\boldsymbol{Y}^\ul \in \mathds{C}^{\Mbsrx \times I_\ul \times V_\ue}$, and whiten them using the corresponding clutter covariances
obtaining $\overline{\boldsymbol{Y}}^\dl$ and $\overline{\boldsymbol{Y}}^\ul$.
We then create the aggregated cube $\overline{\boldsymbol{Y}}=[\overline{\boldsymbol{Y}}^\ul,\overline{\boldsymbol{Y}}^\dl]\in \mathds{C}^{\Mbsrx \times I \times V}$. 
\begin{figure*}[t!]
  \vspace*{-6mm}
  \begin{align}\label{multipage log}
   \log f_{\overline{\yyb}|\eeta}(\overline{\yyb}|\eeta) \hspace{-1mm} =  
   - 
   \hspace{-2mm} 
    \sum_{i,v \in \mathcal{D}} \hspace{-2mm} \left(\overline{\yyb}_{i,v}^\dl\hspace{-1mm} - \overline{\mmu}_{i,v}^\dl \right)^\hermit  \RR_\dl^{-1} \left(\overline{\yyb}_{i,v}^\dl \hspace{-1mm} - \overline{\mmu}_{i,v}^\dl\right) 
    - 
    \sum_{i,v \in \mathcal{U}} \hspace{-2mm} \left(\overline{\yyb}_{i,v}^\ul\hspace{-1mm} - \overline{\mmu}_{i,v}^\ul \right)^\hermit  \RR_\ul^{-1} \left(\overline{\yyb}_{i,v}^\ul \hspace{-1mm} - \overline{\mmu}_{i,v}^\ul\right). 
    \hspace{-1mm}
  \end{align}
\vspace*{-5mm}
  \hrulefill
  \vspace{-1mm}
\end{figure*}
Assuming uncorrelation between $\cc_\dl$ and $\cc_\ul$ and neglecting the constant terms, the log likelihood of $\overline{\yyb} = \veccs{\overline{\YYb}}$ is given in \eqref{multipage log},
where $\RR_\ul= \sqrt[3]{\kappa_{\textrm{cl}}^2 \delta_\textrm{cl}^2}\BB_{\textrm{sp}} + \sigma^2\mathbf{I}_{\Mbsrx}$ 
and $\RR_\dl=\sqrt[3]{\delta_{\textrm{cl}}^2}\BB_{\textrm{sp}} + \sigma^2\mathbf{I}_{\Mbsrx}$.
Here $\overline{\mmu}_{i,v}$ indicates the whitened version of $\overline{\mmu}_{i,v}$, achieved by following the same steps that led to the creation of $\overline{\YYb}$.

\textbf{i) \textit{Clairvoyant Bound}}:
Here, the transmitted symbols are assumed to be always available; they thus do not play a role in the FIM computation. 
The \ac{FIM} w.r.t. $\eeta$ is defined as \cite[Proposition 1]{fodor2024trade} 
\vspace*{-2mm}
    \begin{align}\label{fully det}
        \JJ_{\eeta\eeta}^\textrm{Clair}=&\sum_{\{i,v\} \in \mathcal{D}} 2\Re\left[ 
        \left(\frac{\partial \overline{\mmu}_{i,v}^\dl}{\partial \eeta^\top}\right)^\hermit\RR_{\dl}^{-1} \left(\frac{\partial \overline{\mmu}_{i,v}^\dl}{\partial \eeta^\top}\right)
        \right] + \nonumber\\
        &\sum_{\{i,v\} \in \mathcal{U}} 2\Re\left[ 
        \left(\frac{\partial \overline{\mmu}_{i,v}^\ul}{\partial \eeta^\top}\right)^\hermit\RR_{\ul}^{-1} \left(\frac{\partial \overline{\mmu}_{i,v}^\ul}{\partial \eeta^\top}\right)
        \right].
    \end{align}

\textbf{ii) \textit{Fully Unknown Bound}}:
BS2 now never knows the transmitted symbols, thus treating them as nuisance parameters. Before whitening, each symbol vector affects only its corresponding $\{i,v\}$ bin. Time-frequency whitening may introduce coupling among bins if the whitening matrices are non-diagonal: to retain a tractable formulation, we neglect this whitening-induced coupling and eliminate the unknown symbols independently in each bin.
The  per-bin nuisance parameter vector is
\vspace*{-2mm}
\begin{align}
    \zz_{i,v} = \begin{cases}
        [\Re\{\xx_{i,v}^\top\},\Im\{\xx_{i,v}^\top\}]^\top &   \{i,v\} \in \mathcal{D}  \\
        [\Re\{\xxb_{i,v}^\top\},\Im\{\xxb_{i,v}^\top\}]^\top & \{i,v\} \in \mathcal{U}\\
    \end{cases}
\end{align}
where $\xxb_{i,v}=[\xxb_{1,i,v}^\top,\dots,\xxb_{K,i,v}^\top]^\top$ .
Under the adopted approximation, the global nuisance-parameter FIM is  $\JJ_{\zz\zz}\approx\blkdiag(\JJ_{\zz\zz}^{(i,v)})$.
This approximation substantially reduces the computational complexity by neglecting nuisance coupling, whose size scales as $2KM_{\rm UE}I_{\rm UL}V_{\rm UE}+2I_{\rm DL}(KM_{\rm UE}V_{\rm UE}+V_{\rm S}).$
Assuming $\{i,v\} \in \mathcal{D}$, then the FIM w.r.t. the per-bin augmented parameter vector $\widetilde{\eeta}_{i,v}=[\eeta^\top,\zz_{i,v}^\top]^\top$ is  
$\JJ_{\widetilde{\eeta}\widetilde{\eeta}}^{(i,v)}=
2\Re\left\{ 
         \left[
            \frac{\partial \overline{\mmu}_{i,v}^\dl}{\partial \eeta^\top}~
             \frac{\partial \overline{\mmu}_{i,v}}{\partial \zz_{i,v}^\top} 
        \right]^\hermit
        \hspace{-1mm}
        \RR_\dl^{-1}
        \left[
            \frac{\partial \overline{\mmu}_{i,v}^\dl}{\partial \eeta^\top}~
             \frac{\partial \overline{\mmu}_{i,v}}{\partial \zz_{i,v}^\top} 
        \right]
        \right\}.$
We can now use Schur's complement to isolate the FIM w.r.t. $\eeta$ and subtract the nuisance parameter's contribution: The resulting approximated FIM in the fully-unknown regime is  
\vspace*{-2mm}
\begin{align}
    \JJ_{\eeta\eeta}^\textrm{Nc} =\hspace{-3mm}\sum_{\{i,v\}\in \mathcal{U}\cup \mathcal{D}}\hspace{-3mm}\JJ_{\eeta\eeta}^{(i,v)} -  \Delta\JJ_\eta^{(i,v)},
\end{align}
where $\Delta\JJ_\eta^{(i,v)}=\JJ_{\eeta\zz}^{(i,v)} \left(  \JJ_{\zz\zz}^{(i,v)}\right)^{-1} \JJ_{\zz\eeta}^{(i,v)}$ represents the per-bin information loss induced by the lack of knowledge of the transmitted symbols.

\textbf{iii) \textit{Hybrid Regime}}:
Transmitted symbol knowledge is now available only during UL pilot transmission and within the sensing PRBs, where BS1 transmits pilots.
Let us define the set $\mathcal{P} \subset \mathcal{U}$ containing bins used for pilot transmission and the set  $\mathcal{V}_\textrm{S}$ containing the subcarriers of the sensing PRBs.
Furthermore, the set $\mathcal{S}=\{(i,v) \in \mathcal{D}: v \in \mathcal{V}_S \}$   contains all the bins used for sensing: the set of all the bins where symbol knowledge is available at BS2 is then $\mathcal{K}=\mathcal{P}\cup\mathcal{S}$.
The FIM w.r.t. $\eeta$ in the hybrid regime is defined as 
\vspace*{-2mm}
\begin{align}
    \JJ_{\eeta,\eeta}^\textrm{Hyb}=\hspace{-3mm}\sum_{\{i,v\}\in \mathcal{K} }\hspace{-3mm} \JJ_{\eeta\eeta}^{(i,v)} + \hspace{-3mm}\sum_{\{i,v\} \in (\mathcal{U} \cup \mathcal{D} )\setminus \mathcal{K} }\hspace{-3mm} (\JJ_{\eeta\eeta}^{(i,v)}- \Delta\JJ_{\eeta}^{(i,v)})
\end{align}
    The clairvoyant and fully unknown regimes follow by setting $\mathcal{K}=\mathcal{U} \cup \mathcal{D}$ and $\mathcal{K}=\varnothing$.
\vspace*{-2mm}
\section{Precoding strategy}
Communication and sensing share the transmit waveform and time-frequency resources: UL pilots allow for MMSE channel estimation and provide symbol knowledge to BS2, thus improving its CRB. 
Power allocation impacts both tasks through the power split coefficient $\gamma$: The power budget $P_\textrm{BS}$ of BS1 is divided between communication and sensing as 
$P_{\textrm{BS,c}} = \gamma P_{\textrm{BS}} $ and $P_{\textrm{BS,s}}=(1-\gamma) P_{\textrm{BS}}$.
The power is evenly divided among streams and subcarriers, that is 
\vspace*{-1mm}
\begin{align}
       \rrho_{i,v}=\begin{cases}
    \left[\sqrt{\frac{P_{\textrm{BS,c}}}{V_\textrm{UE}K\Mue}},\dots,\sqrt{\frac{P_{\textrm{BS,c}}}{V_\textrm{UE}K\Mue}}\right] & v \in \mathcal{V}_\textrm{UE},\\
            \sqrt{P_{\textrm{BS,s}}/|\mathcal{V}_\textrm{S}|}  & v \in \mathcal{V}_\textrm{S}.
        \end{cases}
\end{align}
Let us define the set $\mathcal{V}_\textrm{UE}$ containing the subcarriers assigned to the UEs: $\FF_{i,v}$ is then defined as 
\vspace*{-1mm}
\begin{align}
       \FF_{i,v}=\begin{cases}
    [\FF_{i,v,1},\dots,\FF_{i,v,K}] & v \in \mathcal{V}_\textrm{UE},\, i\in \mathcal{D},\\
    \mathbf{a}_\tx(\theta_{\bs,i})  & v \in \mathcal{V}_\textrm{S},\, i\in \mathcal{D}.
        \end{cases}
\end{align}
We adopt \ac{MMSE} precoding for the \acp{UE}: for symbol time $i$ within block $b$, $\FF_{i,v,k}$ is
$    \FF_{i,v,k}\hspace{-1mm} =\hspace{-1mm}\left(\sum_{k=1}^K \alpha_k^2 \left(\widehat{\HH}_{b,k}\widehat{\HH}_{b,k}^\hermit\hspace{-1mm}   + \widetilde{\overline{\boldsymbol{\Xi}}}_{k,b} \hspace{-1mm}\right)+\sigma^2\mathbf{I}_{\Mbstx}\hspace{-1.5mm} \right)^{-1}\hspace{-4mm}\alpha_k\widehat{\HH}_{b,k}$, 
where $ \widetilde{\overline{\boldsymbol{\Xi}}}_{k,b} = \mathbb{E}\left[\widetilde{\HH}_{b,k}\widetilde{\HH}_{b,k}^\hermit \right] $.
BS1 sweeps an angular sector $[  - \Delta\theta/2,  + \Delta\theta/2]$  every frame, thus defining the discretized angle set $\{\theta_{\bs,i}\}_{i=1}^{\tau_\dl}$.
The \acp{UE} evenly spread $P_\textrm{UE}$ among their subcarriers and streams, while their precoding matrices are $\WW_{k,i,v}=\mathbf{I}_{\Mue}$.

\begin{figure}[!htbp]
\begin{center}
   \resizebox{0.48\textwidth}{!}{
%
%
\usetikzlibrary{plotmarks}
\definecolor{mycolor1}{rgb}{0.06600,0.44300,0.74500}%
\definecolor{mycolor2}{rgb}{0.86600,0.32900,0.00000}%
\definecolor{mycolor3}{rgb}{0.92900,0.69400,0.12500}%
\definecolor{mycolor4}{rgb}{0.52100,0.08600,0.81900}%
\definecolor{mycolor5}{rgb}{0.23100,0.66600,0.19600}%
\definecolor{mycolor6}{rgb}{0.18400,0.74500,0.93700}%
\begin{tikzpicture}

\begin{axis}[%
width=0.42\textwidth,
height=1.1793in,
at={(1.888in,0.662in)},
scale only axis,
xmin=0,
xmax=360,
ymin=-300,
ymax=232,
xlabel={$x$ [m]},
ylabel={$y$ [m]},
ytick distance=50,
tick label style={font=\scriptsize},
axis background/.style={fill=white},
axis x line*=bottom,
legend columns=3,
axis y line*=left,
xmajorgrids,
ymajorgrids,
legend style={at={(0.57,0.27)},legend cell align=left, align=left,font=\scriptsize}
]
\addplot [color=mycolor1, line width=2.0pt, only marks, mark size=2.0pt, mark=triangle, mark options={solid, mycolor1}]
  table[row sep=crcr]{%
234.389509671047	-279.334540216866\\
320.51584763567	-166.849989458794\\
354.669697176238	-31.0295777990933\\
348.690036111733	109.941546080199\\
275.743431570397	231.376211730278\\
};
\addlegendentry{UEs}

\addplot [color=mycolor2, line width=2.0pt, only marks, mark size=2.3pt, mark=square, mark options={solid, mycolor2}]
  table[row sep=crcr]{%
0	0\\
};
\addlegendentry{BS 1: $[0,0]$}

\addplot [color=mycolor3, line width=2.0pt, only marks, mark size=2.3pt, mark=square, mark options={solid, mycolor3}]
  table[row sep=crcr]{%
10	30\\
};
\addlegendentry{BS 2: $[10,30]$}

\addplot [color=mycolor4, line width=2.0pt, only marks, mark size=2.0pt, mark=triangle, mark options={solid, rotate=180, mycolor4}]
  table[row sep=crcr]{%
157.6848	157.6848\\
};
\addlegendentry{Tg}

\addplot [color=mycolor5, line width=2.0pt, only marks, mark size=2.5pt, mark=o, mark options={solid, mycolor5}]
  table[row sep=crcr]{%
23.7202376879278	-32.0875865467068\\
22.5809625508804	-32.2332889578661\\
23.9433311571438	-26.3107133019434\\
27.3690000903447	-26.5035858683089\\
26.3812123971537	-27.2671304409952\\
240.463996161255	60.2996419328793\\
240.275605727513	62.1868635268402\\
236.587271931054	64.7959557197263\\
238.185112582204	60.6002034861002\\
237.715390279068	67.5562701477462\\
};
\addlegendentry{UE clusters}

\addplot [color=mycolor6, line width=2.0pt, only marks, mark size=5.0pt, mark=asterisk, mark options={solid, mycolor6}]
  table[row sep=crcr]{%
90	-155.884572681199\\
125.038506682619	-129.481164060957\\
150.960702230176	-98.0350263027049\\
169.144671741464	-61.5636257986204\\
};
\addlegendentry{Clutter patches}

\end{axis}
\end{tikzpicture}%
}
      \vspace{-2mm}
	 \caption{Geometric layout of the simulation scenario.
     }
    \label{geom}
\end{center}
  \vspace{-4mm}
\end{figure}

 {\scriptsize
\begin{table}[!t]\label{table sim}
\vspace*{1.5mm}
\begin{center}
\caption {\centering Simulation parameters default values}
\begin{tabular}{|m{0.5\linewidth} | m{0.4\linewidth}|} 
 \hline
 \textbf{Parameter} & \textbf{value}\\
 \hline\hline 
 BS1 and BS2 antennas ($\Mbstx , \, \Mbsrx$)  & $32,\, 8$\\
 \hline
UEs antennas ($ \Mue$) & $2$\\
 \hline
 UE pathloss ($\alpha_k$)& 3GPP street canyon model \cite{european2018study}\\
 \hline
UE velocities ($\oomega_k$) & $[-1,0]\,\textrm{m/s}$\\
 \hline
 Target velocity and RCS ($\oomega\, \delta^2$) & $[-30,0]\,\textrm{m/s},\,1\,$dB\\
 \hline
N$^\circ$ of PRBs ($N_\textrm{PRB}$)  & $30$\\
  \hline
Subcarriers per block ($V_\textrm{cho}$)  & $20$\\
  \hline
Carrier frequency, Symbol time ($f_c,T$)  & $2$\,GHz, $52\,\mu$s\\
 \hline
 Pilot length ($\tau_\rmp$)  & $K\Mue=10$\\
 \hline
BS and UEs total power ($P_\bs, P_\ue$) & $32,\,10$ dBm\\
    \hline
 UEs and BS noise variance ($\sigma_\bs^2,\, \sigma_k^2$) & $-150\,$dB\\
     \hline
 Clutter texture power ($\delta_\textrm{cl}^2$) & $-120\,$dB\\
 \hline
  UL/DL attenuation constant ($\kappa_\textrm{cl}^2$) & $-20\,$dB\\
 \hline
  Sensing angular sector($\Delta\theta$) & $110^\circ$\\
 \hline
N$^\circ$ of symbols used for sensing ($I$) & $300$\\
 \hline
 Comm./sensing power split ($\gamma$) & $0.5$\\
 \hline
 Pilot sequence repetition factor ($\nu_\rmp$) & $1$\\
 \hline
 \end{tabular}
 \vspace{-5mm}
 \end{center}
\end{table}
}

\section{Numerical Results}
\vspace{-1mm}
The simulations use the network layout in Fig. \ref{geom} and the parameters whose default values are reported in Table I. 
Here $\CC_{\ue,k},\CC_{\bs,k}$ are generated from a clustered power-angle spectrum: each UE is associated with a set of cluster center angles, which scatter rays according to a normal distribution with an angle spread of $1^\circ$\cite{demir2022channel}. Clutter is similarly generated by a superimposition of patches, each having Gaussian angular, Doppler, and delay distributions around its center. 
A complete description of the clutter parameters and covariances can be found in \cite{rivetti2026journal}.

\begin{figure}[t!]
\vspace*{1.5mm}
\centering
\begin{minipage}{\textwidth}
%
%
\definecolor{mycolor1}{rgb}{0.06600,0.44300,0.74500}%
\definecolor{mycolor2}{rgb}{0.86600,0.32900,0.00000}%
\definecolor{mycolor3}{rgb}{0.92900,0.69400,0.12500}%
\definecolor{mycolor4}{rgb}{0.52100,0.08600,0.81900}%
\definecolor{mycolor5}{rgb}{0.00000,1.00000,1.00000}%
\definecolor{mycolor6}{rgb}{0.18400,0.74500,0.93700}%
\definecolor{mycolor7}{rgb}{0.00000,1.00000,1.00000}%
\definecolor{mycolor8}{rgb}{0.12941,0.12941,0.12941}%
\definecolor{green}{rgb}{0.23100,0.66600,0.19600}%
\begin{tikzpicture}

\begin{axis}[%
width=0.41\textwidth,
height=1.01793in,
at={(1.888in,0.962in)},
clip=false,
scale only axis,
ticklabel style={font=\scriptsize},
y label style={
        at={(axis description cs:-0.07,0.5)},
        anchor=south
    },
xmin=-80,
xmax=80,
ymode=log,
ymin=0.02,
ymax=20,
legend columns=3,
yminorticks=true,
ylabel={CRB$(\theta_{\textrm{BS}})\,[\circ]$},
axis background/.style={fill=white},
axis x line*=bottom,
axis y line*=left,
xmajorgrids,
ymajorgrids,
yminorgrids,
legend style={legend cell align=left, align=left, font=\scriptsize}
]
\addplot [color=mycolor1, line width=1.0pt, mark=o, mark options={solid, mycolor1,mark size=1pt},forget plot]
  table[row sep=crcr]{%
-80	0.529082002772216\\
-75.8974358974359	0.379222839604959\\
-71.7948717948718	0.282408497900484\\
-67.6923076923077	0.227979996082613\\
-63.5897435897436	0.175666449933329\\
-59.4871794871795	0.115456921095878\\
-55.3846153846154	0.0837658210995833\\
-51.2820512820513	0.0634207408019161\\
-47.1794871794872	0.0546061978644462\\
-43.0769230769231	0.0448875506859424\\
-38.974358974359	0.0337946047986194\\
-34.8717948717949	0.0291385072498129\\
-30.7692307692308	0.0276334550903641\\
-26.6666666666667	0.0272408492064793\\
-22.5641025641026	0.0262570841669708\\
-18.4615384615385	0.0284777717222166\\
-14.3589743589744	0.0378138219553542\\
-10.2564102564103	0.041312286913364\\
-6.15384615384615	0.0427744335691212\\
-2.05128205128205	0.0424964174261169\\
2.05128205128205	0.0399700627083908\\
6.15384615384615	0.0342022510296413\\
10.2564102564103	0.024842282392297\\
14.3589743589744	0.0242236511166047\\
18.4615384615385	0.0242955181476464\\
22.5641025641026	0.0301186449860301\\
26.6666666666667	0.0383204311225648\\
30.7692307692308	0.0427443985234825\\
34.8717948717949	0.0458876149749728\\
38.974358974359	0.0485340329536136\\
43.0769230769231	0.0510018658409438\\
47.1794871794872	0.0537630830997738\\
51.2820512820513	0.057514373123646\\
55.3846153846154	0.0759939914351877\\
59.4871794871795	0.0993563188500715\\
63.5897435897436	0.155749218797539\\
67.6923076923077	0.190650647746663\\
71.7948717948718	0.241346044944454\\
75.8974358974359	0.307566062259735\\
80	0.416741162546303\\
};

\addplot [color=mycolor2, line width=1.0pt, mark=triangle, mark options={solid, mycolor2,mark size=1pt},forget plot]
  table[row sep=crcr]{%
-80	0.947932401798936\\
-75.8974358974359	0.660700510667975\\
-71.7948717948718	0.49161223077739\\
-67.6923076923077	0.382165038400565\\
-63.5897435897436	0.277246555374118\\
-59.4871794871795	0.152006306098592\\
-55.3846153846154	0.108329325859638\\
-51.2820512820513	0.0777519549415397\\
-47.1794871794872	0.0722992355819146\\
-43.0769230769231	0.0681766185014057\\
-38.974358974359	0.0634297020712917\\
-34.8717948717949	0.060551434631234\\
-30.7692307692308	0.0579795664696583\\
-26.6666666666667	0.0561668437795668\\
-22.5641025641026	0.0545111650031562\\
-18.4615384615385	0.0539831594825585\\
-14.3589743589744	0.0543746489354333\\
-10.2564102564103	0.0548109343105561\\
-6.15384615384615	0.0554412780831016\\
-2.05128205128205	0.0557307865417654\\
2.05128205128205	0.0550953878642539\\
6.15384615384615	0.0534390415631554\\
10.2564102564103	0.0504704429351189\\
14.3589743589744	0.0497950166389104\\
18.4615384615385	0.0489035853889473\\
22.5641025641026	0.050271458493923\\
26.6666666666667	0.051179389167223\\
30.7692307692308	0.0522519693306429\\
34.8717948717949	0.0535966276858291\\
38.974358974359	0.055219314860251\\
43.0769230769231	0.0571226720236347\\
47.1794871794872	0.0596223433946126\\
51.2820512820513	0.0632804270723876\\
55.3846153846154	0.0874414295071533\\
59.4871794871795	0.120451015089875\\
63.5897435897436	0.217344217436771\\
67.6923076923077	0.298130825447825\\
71.7948717948718	0.380272222807431\\
75.8974358974359	0.509865914769619\\
80	0.732252286364064\\
};

\addplot [color=mycolor3, dashed, line width=1.0pt, mark=o, mark options={solid, mycolor3,mark size=1pt},forget plot]
  table[row sep=crcr]{%
-80	2.20774429373237\\
-75.8974358974359	1.71172729088777\\
-71.7948717948718	1.44052900719053\\
-67.6923076923077	1.45674758835646\\
-63.5897435897436	1.49088865621561\\
-59.4871794871795	1.70945871058134\\
-55.3846153846154	1.81648418543925\\
-51.2820512820513	1.85942747252684\\
-47.1794871794872	1.36652875268487\\
-43.0769230769231	0.905254067087245\\
-38.974358974359	0.609957502042686\\
-34.8717948717949	0.502405081948086\\
-30.7692307692308	0.461096269459751\\
-26.6666666666667	0.414323523593715\\
-22.5641025641026	0.362751008109627\\
-18.4615384615385	0.454679638398604\\
-14.3589743589744	0.530644977245267\\
-10.2564102564103	0.27127526119048\\
-6.15384615384615	0.161355277775479\\
-2.05128205128205	0.109316858757783\\
2.05128205128205	0.0786847251151571\\
6.15384615384615	0.0549281832669077\\
10.2564102564103	0.0356156690309773\\
14.3589743589744	0.0325763023437014\\
18.4615384615385	0.0340383726734173\\
22.5641025641026	0.0435115173041513\\
26.6666666666667	0.0690238894255337\\
30.7692307692308	0.0894221126077682\\
34.8717948717949	0.107849163551821\\
38.974358974359	0.125877496788074\\
43.0769230769231	0.140206225554207\\
47.1794871794872	0.150311383711232\\
51.2820512820513	0.168245021060105\\
55.3846153846154	0.205206084937916\\
59.4871794871795	0.250930847909353\\
63.5897435897436	0.350869065694949\\
67.6923076923077	0.45350776080352\\
71.7948717948718	0.679314531831279\\
75.8974358974359	0.970043227102545\\
80	1.45725660676862\\
};

\addplot [color=mycolor4, dashed, line width=1.0pt, mark=triangle, mark options={solid, mycolor4,mark size=1pt},forget plot]
  table[row sep=crcr]{%
-80	20.2939929188776\\
-75.8974358974359	16.4692803521125\\
-71.7948717948718	14.0909913052599\\
-67.6923076923077	14.1043619970715\\
-63.5897435897436	13.132300932173\\
-59.4871794871795	9.89426719405402\\
-55.3846153846154	8.82008780510431\\
-51.2820512820513	7.95288085474015\\
-47.1794871794872	7.07478614948472\\
-43.0769230769231	6.2108189149304\\
-38.974358974359	5.75020737000455\\
-34.8717948717949	5.57753788983229\\
-30.7692307692308	5.26515600266589\\
-26.6666666666667	4.64701088370169\\
-22.5641025641026	4.15092862020207\\
-18.4615384615385	4.64152827968968\\
-14.3589743589744	3.40250910020196\\
-10.2564102564103	1.42744427300063\\
-6.15384615384615	0.780967607607899\\
-2.05128205128205	0.539537171655737\\
2.05128205128205	0.439909644271443\\
6.15384615384615	0.399809908228874\\
10.2564102564103	0.374535787421006\\
14.3589743589744	0.368610683693381\\
18.4615384615385	0.359259776826928\\
22.5641025641026	0.376491663997419\\
26.6666666666667	0.386412386382229\\
30.7692307692308	0.391691250479749\\
34.8717948717949	0.401072302885031\\
38.974358974359	0.419389637208643\\
43.0769230769231	0.440832802173674\\
47.1794871794872	0.465454084861137\\
51.2820512820513	0.502826678912948\\
55.3846153846154	0.697793420386341\\
59.4871794871795	1.18576341524313\\
63.5897435897436	2.62132180061441\\
67.6923076923077	4.46946904115905\\
71.7948717948718	6.47971699424452\\
75.8974358974359	10.0296169791571\\
80	15.2182371426402\\
};

\addplot[area legend, draw=none, fill=red, fill opacity=0.15, forget plot]
table[row sep=crcr] {%
x	y\\
-80	0.02\\
-55	0.02\\
-55	20\\
-80	20\\
}--cycle;

\addplot[area legend, draw=none, fill=mycolor5, fill opacity=0.2, forget plot]
table[row sep=crcr] {%
x	y\\
-53	0.02\\
-15	0.02\\
-15	20\\
-53	20\\
}--cycle;

\addplot[area legend, draw=none, fill=red, fill opacity=0.15, forget plot]
table[row sep=crcr] {%
x	y\\
55	0.02\\
80	0.02\\
80	20\\
55	20\\
}--cycle;

\addlegendimage{empty legend}
\addlegendentry{Noise only:}

\addlegendimage{ mycolor1, line width=1pt, mark=o,mark options={solid, mycolor1,mark size=2pt}}
\addlegendentry{Clair.}

\addlegendimage{ mycolor2, line width=1pt, mark=triangle,mark options={solid, mycolor2,mark size=2pt}}
\addlegendentry{Hyb.}
\addlegendimage{empty legend}
\addlegendentry{With Clutter:}

\addlegendimage{ mycolor3, line width=1pt, dashed, mark=o,mark options={solid, mycolor3,mark size=2pt}}
\addlegendentry{Clair.}

\addlegendimage{ mycolor4, line width=1pt, dashed, mark=triangle,mark options={solid, mycolor4,mark size=2pt}}
\addlegendentry{Hyb.}
\node at (axis description cs:0.5,-0.2) {(a)};


\end{axis}
\end{tikzpicture}%
\
\end{minipage}
\begin{minipage}{\textwidth}
%
%
\definecolor{mycolor1}{rgb}{0.06600,0.44300,0.74500}%
\definecolor{mycolor2}{rgb}{0.86600,0.32900,0.00000}%
\definecolor{mycolor3}{rgb}{0.92900,0.69400,0.12500}%
\definecolor{mycolor4}{rgb}{0.52100,0.08600,0.81900}%
\definecolor{mycolor5}{rgb}{0.00000,1.00000,1.00000}%
\definecolor{mycolor6}{rgb}{0.12941,0.12941,0.12941}%
\definecolor{green}{rgb}{0.23100,0.66600,0.19600}%
\begin{tikzpicture}

\begin{axis}[%
width=0.41\textwidth,
height=1.01793in,
clip=false,
at={(1.888in,0.962in)},
scale only axis,
legend columns=3,
tick label style={font=\scriptsize},
xmin=-80,
xmax=80,
y label style={
        at={(axis description cs:-0.07,0.5)},
        anchor=south
    },
x label style={
        at={(axis description cs:0.5,-0.3)},
        anchor=south
    },
xlabel style={font=\color{mycolor6}},
xlabel={$ \theta_{\textrm{BS}} [\circ]$},
ymode=log,
ymin=0.2,
ymax=200,
yminorticks=true,
ylabel style={font=\color{mycolor6}},
ylabel={CRB$(\tau_{\textrm{BS}})$\, [m]},
yticklabel style={xshift=1pt}
axis background/.style={fill=white},
axis x line*=bottom,
axis y line*=left,
xmajorgrids,
ymajorgrids,
yminorgrids,
legend style={legend cell align=left, align=left, font=\scriptsize}
]
\addplot [color=mycolor1, line width=1.0pt, mark=o, mark options={solid, mycolor1, mark size=1.5pt}, forget plot]
  table[row sep=crcr]{%
-80	1.32293511734669\\
-75.8974358974359	1.34230916133325\\
-71.7948717948718	1.32101130225177\\
-67.6923076923077	1.15344781116875\\
-63.5897435897436	1.1234718694773\\
-59.4871794871795	0.797811683085434\\
-55.3846153846154	0.524919319255172\\
-51.2820512820513	0.449837186217758\\
-47.1794871794872	0.421542880444906\\
-43.0769230769231	0.37976635017432\\
-38.974358974359	0.247219086158184\\
-34.8717948717949	0.240408473320125\\
-30.7692307692308	0.229129442262675\\
-26.6666666666667	0.240965044322291\\
-22.5641025641026	0.237708833850832\\
-18.4615384615385	0.280797991704619\\
-14.3589743589744	0.378059132260072\\
-10.2564102564103	0.433813869085031\\
-6.15384615384615	0.461784719368089\\
-2.05128205128205	0.464406023476443\\
2.05128205128205	0.436127219817443\\
6.15384615384615	0.366938775387451\\
10.2564102564103	0.227677963521242\\
14.3589743589744	0.248959832006099\\
18.4615384615385	0.211012285996725\\
22.5641025641026	0.329844030122861\\
26.6666666666667	0.397236881809191\\
30.7692307692308	0.433578505310943\\
34.8717948717949	0.449461105990106\\
38.974358974359	0.454228045926761\\
43.0769230769231	0.455879376467676\\
47.1794871794872	0.457088754072925\\
51.2820512820513	0.450124642287593\\
55.3846153846154	0.474849364765675\\
59.4871794871795	0.751255797571367\\
63.5897435897436	0.956247540699109\\
67.6923076923077	0.990145107337233\\
71.7948717948718	1.03503196127816\\
75.8974358974359	1.06998329778836\\
80	1.06265223770502\\
};

\addplot [color=mycolor2, line width=1.0pt, mark=triangle, mark options={solid, mycolor2,mark size=1.5pt}, forget plot]
  table[row sep=crcr]{%
-80	3.06941421156245\\
-75.8974358974359	3.13713460945519\\
-71.7948717948718	3.04263001608433\\
-67.6923076923077	2.63589595831474\\
-63.5897435897436	2.47983148874033\\
-59.4871794871795	1.46893672294387\\
-55.3846153846154	0.739653831664222\\
-51.2820512820513	0.653179945565935\\
-47.1794871794872	0.662924192780101\\
-43.0769230769231	0.673515984099562\\
-38.974358974359	0.668908265903061\\
-34.8717948717949	0.677495266983024\\
-30.7692307692308	0.683384339378364\\
-26.6666666666667	0.692048079521654\\
-22.5641025641026	0.695097207833644\\
-18.4615384615385	0.704942720922371\\
-14.3589743589744	0.715628053589314\\
-10.2564102564103	0.714190673200849\\
-6.15384615384615	0.708471374177046\\
-2.05128205128205	0.701381250230981\\
2.05128205128205	0.693695991051462\\
6.15384615384615	0.684015999468131\\
10.2564102564103	0.660311632015076\\
14.3589743589744	0.655200192093877\\
18.4615384615385	0.63797228899219\\
22.5641025641026	0.643245051452152\\
26.6666666666667	0.632968679089515\\
30.7692307692308	0.617765813011289\\
34.8717948717949	0.600418831741753\\
38.974358974359	0.582182608719069\\
43.0769230769231	0.563985586431413\\
47.1794871794872	0.546249392297626\\
51.2820512820513	0.530991373212571\\
55.3846153846154	0.592102641023621\\
59.4871794871795	1.18714176061208\\
63.5897435897436	1.97970050252409\\
67.6923076923077	2.09228745206907\\
71.7948717948718	2.38325776237324\\
75.8974358974359	2.46211404917457\\
80	2.41480938459784\\
};

\addplot [color=mycolor3, dashed, line width=1.0pt, mark=o, mark options={solid, mycolor3,mark size=1.5pt}, forget plot]
  table[row sep=crcr]{%
-80	6.85531265640782\\
-75.8974358974359	7.6094288290537\\
-71.7948717948718	8.63866986743033\\
-67.6923076923077	9.64414219653876\\
-63.5897435897436	12.3474175781888\\
-59.4871794871795	14.9240940096447\\
-55.3846153846154	17.4433440487461\\
-51.2820512820513	18.0949197712703\\
-47.1794871794872	15.5454013343824\\
-43.0769230769231	12.1750601186416\\
-38.974358974359	6.3899958437596\\
-34.8717948717949	5.82949634805239\\
-30.7692307692308	5.22160118129756\\
-26.6666666666667	5.0465811802646\\
-22.5641025641026	4.50218898910561\\
-18.4615384615385	6.64641778073653\\
-14.3589743589744	7.94779231487141\\
-10.2564102564103	4.12953559167535\\
-6.15384615384615	2.50958811674538\\
-2.05128205128205	1.78592630823901\\
2.05128205128205	1.37517436530861\\
6.15384615384615	1.01921086121367\\
10.2564102564103	0.451973748199019\\
14.3589743589744	0.516734822211773\\
18.4615384615385	0.388708237272724\\
22.5641025641026	0.832566036350942\\
26.6666666666667	1.09748787178019\\
30.7692307692308	1.23992084137585\\
34.8717948717949	1.3230514811516\\
38.974358974359	1.38714812920425\\
43.0769230769231	1.43046518635634\\
47.1794871794872	1.45795167658736\\
51.2820512820513	1.46972472912877\\
55.3846153846154	1.58461712801018\\
59.4871794871795	2.37572973746731\\
63.5897435897436	2.89088090495522\\
67.6923076923077	3.3664124619142\\
71.7948717948718	3.79665216524462\\
75.8974358974359	4.46364174421085\\
80	4.92476593986104\\
};

\addplot [color=mycolor4, dashed, line width=1.0pt, mark=triangle, mark options={solid, mycolor4,mark size=1.5pt}, forget plot]
  table[row sep=crcr]{%
-80	73.2233018028132\\
-75.8974358974359	80.6561373931934\\
-71.7948717948718	98.9838829230949\\
-67.6923076923077	104.076505454418\\
-63.5897435897436	136.167600616595\\
-59.4871794871795	94.1767610266603\\
-55.3846153846154	67.4302674207675\\
-51.2820512820513	68.7084144690347\\
-47.1794871794872	66.804078921356\\
-43.0769230769231	63.8166950183347\\
-38.974358974359	63.4704966624933\\
-34.8717948717949	65.2318896347481\\
-30.7692307692308	64.4488661251739\\
-26.6666666666667	58.9032705558328\\
-22.5641025641026	54.3841629319191\\
-18.4615384615385	62.1524935969994\\
-14.3589743589744	45.8868407302842\\
-10.2564102564103	19.2683458773082\\
-6.15384615384615	10.4903309805674\\
-2.05128205128205	7.20849672426528\\
2.05128205128205	5.8817841913889\\
6.15384615384615	5.38809686246269\\
10.2564102564103	5.102269782146\\
14.3589743589744	5.00323971808809\\
18.4615384615385	4.80000410593345\\
22.5641025641026	4.86947258204339\\
26.6666666666667	4.80923332531312\\
30.7692307692308	4.67399564089446\\
34.8717948717949	4.571292090334\\
38.974358974359	4.53117546078562\\
43.0769230769231	4.45188460811467\\
47.1794871794872	4.30040201198024\\
51.2820512820513	4.22121958951426\\
55.3846153846154	5.08545911890338\\
59.4871794871795	11.2082149667823\\
63.5897435897436	26.6890466550306\\
67.6923076923077	32.6465402504901\\
71.7948717948718	45.5840106950047\\
75.8974358974359	49.3258180718244\\
80	54.1529645991017\\
};

\addplot[area legend, draw=none, fill=red, fill opacity=0.15, forget plot]
table[row sep=crcr] {%
x	y\\
-80	0.2\\
-55	0.2\\
-55	200\\
-80	200\\
}--cycle;

\addplot[area legend, draw=none, fill=mycolor5, fill opacity=0.2, forget plot]
table[row sep=crcr] {%
x	y\\
-53	0.2\\
-15	0.2\\
-15	200\\
-53	200\\
}--cycle;

\addplot[area legend, draw=none, fill=red, fill opacity=0.15, forget plot]
table[row sep=crcr] {%
x	y\\
55	0.2\\
80	0.2\\
80	200\\
55	200\\
}--cycle;






\node at (axis description cs:0.5,-0.35) {(b)};


\end{axis}
\end{tikzpicture}%
\end{minipage}
\vspace{-3mm}
\caption{CRB regimes of $\theta_\textrm{BS}$ (a), and $\tau_\textrm{BS}$ (b) as a function of  $\theta_\textrm{BS}$. The red areas represent the angles with $|\theta_\textrm{BS}|>\Delta\theta/2$, while the cyan one is the angular sector occupied by the clutter patches. The "noise only" regime is obtained by setting $\RR_\ul= \RR_\dl= \sigma^2_\bs\mathbf{I}_{\Mbsrx}$ and  using the un-whitened $\mmu_{i,v}$. The fully unknown bound is here not reported for the sake of visual clarity: it has a noise-like nature, fluctuating around $10^4$ for $\theta_\textrm{BS}$ and $10^5$ for $\tau_\textrm{BS}$.  } 
\label{subfig theta}
\end{figure}


\begin{figure}[t!]
\centering

\begin{minipage}{\textwidth}
%
%
\definecolor{mycolor1}{rgb}{0.06600,0.44300,0.74500}%
\definecolor{mycolor2}{rgb}{0.86600,0.32900,0.00000}%
\definecolor{mycolor3}{rgb}{0.92900,0.69400,0.12500}%
\definecolor{mycolor4}{rgb}{0.52100,0.08600,0.81900}%
\definecolor{mycolor5}{rgb}{0.23100,0.66600,0.19600}%
\definecolor{mycolor6}{rgb}{0.18400,0.74500,0.93700}%
\definecolor{mycolor7}{rgb}{0.12941,0.12941,0.12941}%
\begin{tikzpicture}[spy using outlines={rectangle, magnification=2.5,connect spies}]

\begin{axis}[%
width=0.42\textwidth,
height=1.2793in,
clip=false,
at={(1.71in,0.724in)},
scale only axis,
xmin=0,
xmax=100,
legend columns=3,
xlabel style={font=\color{mycolor7}},
xlabel={$ \gamma [\%]$},
ymode=log,
ymin=0.3,
ymax=200.9939931996168,
yminorticks=true,
ylabel style={font=\color{mycolor7}},
ylabel={CRB$(\tau_{\textrm{BS}})$\, [m]},
axis background/.style={fill=white},
axis x line*=bottom,
axis y line*=left,
xmajorgrids,
ymajorgrids,
yminorgrids,
tick label style={font=\scriptsize},
y label style={
        at={(axis description cs:-0.06,0.5)},
        anchor=south
    },
x label style={
        at={(axis description cs:0.5,-0.2)},
        anchor=south
    },
legend style={at={(0.15,0.604)}, anchor=south west, legend cell align=left, align=left, font=\scriptsize}
]


\addlegendimage{ mycolor4, line width=1pt}
\addlegendentry{$\theta_\textrm{BS}=-41^\circ$}

\addlegendimage{ mycolor1, line width=1pt}
\addlegendentry{$\theta_\textrm{BS}=45^\circ$}

\addlegendimage{ mycolor5, line width=1pt}
\addlegendentry{$\theta_\textrm{BS}=15^\circ$}

 \node[
    draw=black,
    fill=white,
    font=\scriptsize,
    anchor=north west,
    align=left,
    inner sep=4pt
] at (rel axis cs:0.05,1) {
\raisebox{0.3ex}{\tikz \draw[thick,dashed,mark=triangle,mark indices={2},mark options={solid}] plot coordinates {(0,0) (4,0) (8,0)};}~Hyb.\\
\raisebox{0.3ex}{\tikz \draw[thick, mark=o,mark indices={2}, mark options={solid}] plot coordinates {(0,0) (4,0) (8,0)};}~Clair.
};



\addplot [color=mycolor1, line width=1.0pt, mark=o, mark options={solid, mycolor1, mark size=1.5pt}, forget plot]
  table[row sep=crcr]{%
1	2.76613246530311\\
6.15789473684211	2.10062845448723\\
11.3157894736842	1.83689069455129\\
16.4736842105263	1.69478609079207\\
21.6315789473684	1.60788448326764\\
26.7894736842105	1.55149835486198\\
31.9473684210526	1.51417625834084\\
37.1052631578947	1.48984145024247\\
42.2631578947368	1.47496863288543\\
47.4210526315789	1.46738473075957\\
52.5789473684211	1.4656961630922\\
57.7368421052632	1.46899080161812\\
62.8947368421053	1.4766726992829\\
68.0526315789474	1.48836626125807\\
73.2105263157895	1.50385931911045\\
78.3684210526316	1.52306950102807\\
83.5263157894737	1.54602562199631\\
88.6842105263158	1.5728596699837\\
93.8421052631579	1.60380701148339\\
99	1.63921099686803\\
};


\addplot [color=mycolor1, line width=1.0pt, mark=triangle, mark options={solid, mycolor1, mark size=1.5pt}, forget plot,dashed]
  table[row sep=crcr]{%
1	3.167608314151\\
6.15789473684211	3.24680036918662\\
11.3157894736842	3.3333204524993\\
16.4736842105263	3.42825153354767\\
21.6315789473684	3.53291218837657\\
26.7894736842105	3.64894273972264\\
31.9473684210526	3.77841387093651\\
37.1052631578947	3.92398193050603\\
42.2631578947368	4.08911851552121\\
47.4210526315789	4.27846074961341\\
52.5789473684211	4.49836530390661\\
57.7368421052632	4.75782286689366\\
62.8947368421053	5.0700475773988\\
68.0526315789474	5.45542204226403\\
73.2105263157895	5.94741722282484\\
78.3684210526316	6.6058466696925\\
83.5263157894737	7.55136498790371\\
88.6842105263158	9.07839210430873\\
93.8421052631579	12.2123838157834\\
99	28.556286192581\\
};
 

\addplot [color=mycolor5, line width=1.0pt, mark=o, mark options={solid, mycolor5, mark size=1.5pt}, forget plot]
  table[row sep=crcr]{%
1	1.6778166339231\\
6.15789473684211	1.0767589057233\\
11.3157894736842	0.905331368756625\\
16.4736842105263	0.801588545731343\\
21.6315789473684	0.72830779559776\\
26.7894736842105	0.672548232953202\\
31.9473684210526	0.628131063560892\\
37.1052631578947	0.591607099367492\\
42.2631578947368	0.560857151843576\\
47.4210526315789	0.534490160990336\\
52.5789473684211	0.51154715818692\\
57.7368421052632	0.491341392807876\\
62.8947368421053	0.473365777447712\\
68.0526315789474	0.457236323127661\\
73.2105263157895	0.442656027412257\\
78.3684210526316	0.429390972672846\\
83.5263157894737	0.417254015688187\\
88.6842105263158	0.406093357260777\\
93.8421052631579	0.395784318594968\\
99	0.386222967391026\\
};

\addplot [color=mycolor5, line width=1.0pt, mark=triangle, mark options={solid, mycolor5, mark size=1.5pt}, forget plot, dashed]
  table[row sep=crcr]{%
1	3.7043547551365\\
6.15789473684211	3.76488553370543\\
11.3157894736842	3.85399328444909\\
16.4736842105263	3.9574283909138\\
21.6315789473684	4.07320581557514\\
26.7894736842105	4.20181660933082\\
31.9473684210526	4.34479709270443\\
37.1052631578947	4.50447015858339\\
42.2631578947368	4.68401091976493\\
47.4210526315789	4.88769314751406\\
52.5789473684211	5.12133195098924\\
57.7368421052632	5.39302795645702\\
62.8947368421053	5.71444740046802\\
68.0526315789474	6.10314707519646\\
73.2105263157895	6.58713803683804\\
78.3684210526316	7.21485297675178\\
83.5263157894737	8.08050547841257\\
88.6842105263158	9.40591791589198\\
93.8421052631579	11.9489199827185\\
99	25.3349920854324\\
};

\addplot [color=mycolor4, line width=1.0pt, mark=o, mark options={solid, mycolor4, mark size=1.5pt}, forget plot]
  table[row sep=crcr]{%
1	26.345831089667\\
6.15789473684211	16.930624401184\\
11.3157894736842	14.4664315339697\\
16.4736842105263	13.0001909848734\\
21.6315789473684	11.9590155926511\\
26.7894736842105	11.1588795733863\\
31.9473684210526	10.5148458371273\\
37.1052631578947	9.98004842121566\\
42.2631578947368	9.5257339049408\\
47.4210526315789	9.132968431319\\
52.5789473684211	8.78863227118343\\
57.7368421052632	8.48327484825299\\
62.8947368421053	8.20987623374049\\
68.0526315789474	7.96308968638434\\
73.2105263157895	7.73875680295892\\
78.3684210526316	7.53358543587021\\
83.5263157894737	7.34492902440822\\
88.6842105263158	7.1706314065312\\
93.8421052631579	7.00891525470969\\
99	6.85830102197728\\
};


\addplot [color=mycolor4, line width=1.0pt, mark=triangle, mark options={solid, mycolor4, mark size=1.5pt}, forget plot,dashed]
  table[row sep=crcr]{%
1	45.5562872700442\\
6.15789473684211	46.7562783999213\\
11.3157894736842	48.0588823816651\\
16.4736842105263	49.4784575710277\\
21.6315789473684	51.0332953992924\\
26.7894736842105	52.7461495266584\\
31.9473684210526	54.6456107636802\\
37.1052631578947	56.7681419797578\\
42.2631578947368	59.1610847508295\\
47.4210526315789	61.8872151690026\\
52.5789473684211	65.0318828159297\\
57.7368421052632	68.7146633732431\\
62.8947368421053	73.1093451131064\\
68.0526315789474	78.4803779995804\\
73.2105263157895	85.2547389621288\\
78.3684210526316	94.179013666928\\
83.5263157894737	106.715915873632\\
88.6842105263158	126.29112204869\\
93.8421052631579	164.12972434265\\
99	341.763200608196\\
};

\end{axis}
\end{tikzpicture}%
\end{minipage}
\vspace{-2mm}
\caption{Clutter affected CRB for different target positions, as a function of the power trade-off parameter $\gamma$} 
\vspace{-4mm}
\label{fig gamma}
\end{figure}




\begin{figure}[t!]
\centering
\begin{minipage}{\textwidth}
\input{images/provisional_results/SE/SE_subfig_avg_a}
\end{minipage}
\vspace{0mm}
\begin{minipage}{\textwidth}
\vspace{-2mm}
\input{images/provisional_results/SE/SE_subfig_avg_b}
\end{minipage}
\vspace{0mm}
\begin{minipage}{\textwidth}
 \vspace{-2mm}
\input{images/provisional_results/SE/SE_subfig_avg_c}
\end{minipage}
\vspace{-2mm}
\caption{\ac{CDF} of the average UL sum SE as a function of $P_\ue$ and $\nu_\rmp$ (a), UE's velocity magnitude $\norm{\oomega}_k$ (b) and $\nu_\rmp$ (c). These \acp{CDF} are obtained with $1000$ Monte Carlo realizations. } 
\vspace{-3mm}
\label{subfig SE}
\end{figure}

\textbf{i) \textit{Sensing Accuracy}}:
The CRBs are defined as
$\sqrt{\left[\JJ\right]_{l,l}^{-1}}$, with $l$ being the parameter's position in $\eeta$.
In this section, unless otherwise specified, we assume that $\tau_\textrm{c}=60$, $\tau_\textrm{DL}=30$, $N_\textrm{PRB}^\textrm{UE}=15$.
The reported CRBs are obtained by inverting the Monte Carlo-averaged FIMs over 100 independent realizations of the transmit waveform across the entire sensing window.
We applied the per-bin nuisance approximation described in Section IV.
Reduced-scale comparisons with global nuisance elimination indicate that this approximation is optimistic in the considered setup. The per-bin approach underestimates the hybrid CRB by roughly $63\%$ for both $\theta_\textrm{BS}$ and $\tau_\textrm{BS}$.
Fig. \ref{subfig theta} shows the \acp{CRB} evolution as a function of $\theta_\bs$. We see a clear gain when the target ``leaves" the clutter-affected area, while all \acp{CRB} soar when the target leaves $\Delta\theta$. Interestingly, within $\Delta \theta$, the "noise-only" regime seems to be less affected by imperfect symbol knowledge. Furthermore, the hybrid-clairvoyant gap increases when the target is aligned with a set of \ac{UE}-communication clusters, with the clairvoyant curves having a dip at $\theta_\bs=-30^\circ$ and $\theta_\bs=15^\circ$. In Fig. \ref{fig gamma}, we analyze the impact of  $\gamma$. 
The hybrid \acp{CRB} soar as more power is allocated to communications, which is expected given the lack of knowledge on the communication symbols.
On the other hand, the clairvoyant CRB benefits from a higher communication power, as the CRB decreases with $\gamma$.
Interestingly, when the target is at $\theta_\textrm{BS}=45^\circ$, the clairvoyant \ac{CRB} exhibits a slightly convex behavior, reaching an optimum at $\gamma\approx 52\%$.

\textbf{ii) \textit{Aging and Spectral Efficiency}}:
In Fig.~\ref{subfig SE}, unless otherwise specified, we assume that  $\tau_\textrm{c}=30$, $\tau_\textrm{DL}=15$, $p=2$, and $N_\textrm{PRB}^\textrm{UE}=2$.
We here show the \ac{CDF} of the \acp{UE} \ac{UL} \ac{SE} as a function of key network parameters.
Fig. \ref{subfig SE}a shows the impact of pilot subcarrier occupation, $\nu_\rmp$, for different $P_\ue$. 
Here, increasing $\nu_\rmp$ beyond $1$ is beneficial only at low SNR ($P_\textrm{UE}=10$\,dBm), while it becomes harmful as $P_\textrm{UE}$ increases.
This is explained by the fact that increasing $\nu_\rmp$ increases the \acp{UE}' \ac{SNR}, but reduces the fraction of resources available for data transmission: when in an \ac{SNR}-limited regime, the \ac{SNR} gain dominates; conversely, at higher \ac{SNR}s, the pre-log penalty becomes the limiting factor.
Then, Fig. \ref{subfig SE}b shows the impact of UE mobility and how that is inversely proportional to the gain brought by using $p$ past pilot observations during channel estimation. Indeed, the gap between the CDF curves corresponding to $p=1$ and $p=5$ is substantially lower when the UEs' velocity increases from $1$ to $5$\,m/s. 
Note that the curve with $p=0$ corresponds to a standard block-fading channel estimator, which uses only the present pilot observation as the previous ones are assumed to be uncorrelated.
Interestingly, a similar trend is observed
when the UEs repeat $\beth_k$ across $\nu_\rmp=5$ subcarriers, in Fig. \ref{subfig SE}c, the dotted curves present a much lower sensitivity to $p$, indicating that pilot repetition across frequency diminishes the advantage of known channel temporal correlation.

\section{conclusion}
This paper assesses the impact of transmit-symbol uncertainty and clutter on bistatic ISAC sensing accuracy through a CRB analysis.
At the same time, the inevitable communication channel correlation is exploited through the definition of a correlation-aware channel estimation.
Even under the adopted per-bin approximation, which reduced-scale comparisons indicate to be optimistic, symbol uncertainty significantly degrades estimation accuracy. Therefore, accounting for the full nuisance coupling is expected to reinforce the impact of symbol uncertainty. The magnitude of the degradation depends on target geometry and clutter conditions. Symbol knowledge is shown to impact the choice of the power-split parameter $\gamma$. Furthermore, pilot design and channel temporal correlation can be jointly exploited to realize substantial SE gain over conventional block-fading schemes. These results provide practical design insights for distributed ISAC systems.
\vspace{-5mm}



\bibliographystyle{IEEEtran}
\bibliography{IEEEabrv,auxiliary/biblio,JCASv10}
\end{document}